\documentclass{webofc}
\usepackage[varg]{txfonts}   
\newcommand{\be}{\begin{equation}}
\newcommand{\ee}{\end{equation}}
\newcommand{\bea}{\begin{eqnarray}}
\newcommand{\eea}{\end{eqnarray}}
\begin{document}
\title{Forbidden Channels and SIMP Dark Matter}
%
%

\author{\firstname{Soo-Min} \lastname{Choi}\inst{1}\fnsep\thanks{\email{soominchoi90@gmail.com}} \and
        \firstname{Yoo-Jin} \lastname{Kang}\inst{1,2}\fnsep\thanks{\email{yoojinkang91@gmail.com}} \and
        \firstname{Hyun Min} \lastname{Lee}\inst{1}\fnsep\thanks{\email{hmlee71@gmail.com}}
}

\institute{Department of Physics, Chung-Ang University, Seoul 06974, Korea
\and
           Center for Theoretical Physics of the Universe, Institute for Basic Science, Daejeon 34051, Korea
 }

\abstract{%
  In this review, we focus on dark matter production from thermal freeze-out with forbidden channels and SIMP processes. We show that forbidden channels can be dominant to produce dark matter depending on the dark photon and / or dark Higgs mass compared to SIMP.
}
\maketitle

\begin{center}
{\it \small Prepared for the proceedings of the 13th International Conference on Gravitation,\\
Ewha Womans University, Korea, 3-7 July 2017. }
\end{center}

\section{Introduction}
\label{intro}

 In the Universe, there are diverse evidences for dark matter such as galaxy rotation curves, gravitational lensing and Cosmic Microwave Background (CMB). Especially, CMB from Planck satellite data infer that the averaged relic density of dark matter is $\Omega_{\rm DM}h^2=0.1198\pm 0.0015$ \cite{Planck}.
 
 Various dark matter candidates have been introduced including Weakly Interacting Massive Particles (WIMPs). WIMPs are well-motivated but carry too small self-interaction to explain small-scale problems. Small-scale problems come from difference between numerical simulations with collisionless dark matter and observations from galaxies. They include core-cusp problem, missing satellite problem and too-big-to-fail problem. Solutions to small-scale problems require dark matter as self-interaction strong as $\sigma_{\rm self}/m_{\rm DM}=0.1\ -\ 10{\rm cm^2/g}$ \cite{small1, small2}.

In this sence, dark matter annihilating with forbidden channels can be important. In the forbidden channels, masses of final states are larger than dark matter masses. These channels are forbidden at zero temperature but open at high temperature in the early Universe \cite{forb}. The forbidden channel needs dark matter with sizable self-interaction, since they have Boltzmann suppression factor. It is discussed in section 3.1 in this review. 
 
 Strongly Interacting Massive Particles (SIMP) \cite{simp1} have started drawing some attention recently because they can allow large self-interacting of light dark matter. The relic density of SIMP is determined by $3 \rightarrow 2$  self annihilations in the thermal freeze-out \cite{simp2}. However, the Boltzmann suppression factor makes SIMP be on the verge of violating unitarity or perturbativity. Also it makes SIMP be in a tension with the bounds from Bullet cluster and halo shapes in the most of the parameter space \cite{simp3}.  

 In this review, we consider a complex scalar dark matter, dark photon and dark Higgs. We discuss dark matter can be produced from thermal freeze-out with forbidden channels and/or SIMP processes. In forbidden channels, annihilation cross section can be suppressed by a Boltzmann factor which depends on mass difference, $\Delta_i=(m_i-m_{\rm DM})/m_{\rm DM}$, between dark matter and dark photon or dark Higgs. We suggest a model with gauged $Z_3$ symmetry for forbidden and SIMP dark matter. Specific range of mass differences allow dark matter to have a smaller self-interaction being compatible with the relic density, as compared to the SIMP case. In our model, the self-coupling contributes to both SIMP and forbidden channels. 
 
\section{Model for self-interacting dark matter}
\label{model}

 We  consider a complex scalar dark matter $\chi$ and dark Higgs $\phi$ in dark local $U(1)_d$ model with charge $q_\chi=+1$ and $q_\phi=+3$ each. Local $U(1)_d$ symmetry is spontaneously broken to $Z_3$ which makes dark matter stable. Then, the dark Higgs $\phi$ has non-zero vacuum expectation value $\langle \phi\rangle=v_d/\sqrt{2}$ \cite{z3a, z3b}.
 
  The Lagrangian for SM singlet scalars, $\chi$ and $\phi$, and the SM Higgs doublet $H$, is given by
\bea
{\cal L}=-\frac{1}{4}V_{\mu \nu}V^{\mu\nu}-\frac{1}{2}\sin{\xi}V_{\mu\nu}B^{\mu\nu}+|D_\mu \phi |^2+|D_\mu \chi |^2+|D_\mu H |^2-V(\phi,\chi,H)
\eea
where $V_{\mu\nu}=\partial_\mu V_\nu -\partial_\nu V_\mu$, $D_\mu \phi=(\partial_\mu-iq_\phi g_d V_\mu)\phi$, $D_\mu \chi=(\partial_\mu-iq_\chi g_d V_\mu)\chi$, with dark gauge coupling $g_d$, and $D_\mu H=(\partial_\mu-ig' Y_H B_\mu-\frac{1}{2}igT^a W_\mu^a)H$. The second term means mixing between dark photon $Z'$ and SM gauge boson $Z$. This  $V(\phi,\chi,H)=V_{\rm DM}+V_{\rm SM}$ is the scalar potential and $V_{\rm SM}$ is the Standard Model Higgs potential. $V_{\rm DM}$ is
\bea
V_{\rm DM}&=&-m_\phi^2 |\phi |^2 +m_\chi^2 |\chi |^2 +\lambda_\phi |\phi |^4+\lambda_\chi |\chi |^4 +\lambda_{\phi \chi}|\phi |^2|\chi |^2+\bigg(\frac{\sqrt{2}}{3!}\kappa\phi^\dagger \chi^3 + {\rm h.c.} \bigg) \nonumber \\
&&+ \lambda_{\phi H}|\phi |^2|H |^2+\lambda_{\chi H}|\chi |^2 |H |^2.
\eea
After the dark Higgs is expanded as $\phi=(v_d+h_d)/\sqrt{2}$, a triple coupling for $\chi$ is obtained.  This coupling makes dark matter annihilate by $3\rightarrow 2$ process.

\section{Thermal freeze-out}
\label{freeze}

We discuss the thermal production of light dark matter from process in combination with dark Higgs quartic coupling and $Z'$ gauge coupling with the 2$\rightarrow$2 forbidden channels and the 3$\rightarrow$2 annihilation SIMP both by dark matter self-interactions. From that, we can find which process would be dominant process to determine the relic density depending on the mass difference between dark matter and dark photon and/or dark Higgs. Moreover, depending on dark photon and dark Higgs masses, annihilation into a pair of the SM particles also can be dominant. The SM-annihilation is discussed in the original paper \cite{orig}. In this review, we focus on comparison between forbidden channels and SIMP channels.  

\subsection{Forbidden channels}
\label{forb}

Although the dark photon and/or dark Higgs are heavier than dark matter, they can still contribute to the relic density for dark matter through the forbidden channels. We consider the case where dark photon and / or dark Higgs are light. We assume that dark photon and/or dark Higgs are in kinetic equilibrium during freeze-out process.

 If we consider the annihilation cross section for forbidden channels for example $\chi \chi^*\rightarrow Z'Z'$ and $\chi \chi^* \rightarrow Z' \chi^*$, the detailed balance condition at high temperature are 
\bea
\langle \sigma v \rangle_{\chi \chi^* \rightarrow Z' Z'}&=&\frac{4(n_{Z'}^{\rm eq})^2}{(n_{\rm DM}^{\rm eq})^2} \langle \sigma v \rangle_{Z' Z'\rightarrow \chi \chi^* } \nonumber \\
&=& 9(1+\Delta_{Z'})^3 e^{-2\Delta_{Z'}x}\langle \sigma v \rangle_{Z' Z'\rightarrow \chi \chi^* }
\eea and similarly
\bea
\langle \sigma v \rangle_{\chi \chi \rightarrow Z' \chi^*}&=&3(1+\Delta_{Z'})^3 e^{-2\Delta_{Z'}x} \langle \sigma v \rangle_{Z' \chi^*\rightarrow \chi \chi } 
\eea with $\Delta_{Z'} \equiv (m_{Z'}-m_\chi)/ m_{\chi}$ and $x=m_{\chi}/T$.

From (4) and (5), we can see that the annihilation cross sections for forbidden channels depend on $e^{-\Delta_{Z'}x}$. Therefore, in the case with $m_\chi < m_{Z'} \ll m_{h_1}$, dark Higgs can't contribute to the relic density for dark matter. Similarly, in the case with $m_\chi < m_{h_1} \ll m_{Z'}$, dark photon can't contribute to the relic density for dark matter. However, the both dark photon and dark Higgs can contribute to the forbidden channels at the same time in the case with $m_\chi < m_{Z'} \sim m_{h_1}$. The first case are discussed below. The other casse are discussed with the first case in the original paper \cite{orig}. 

In the case with $m_\chi < m_{Z'} \ll m_{h_1}$, the Boltzmann equation is approximated to 
\bea
\frac{dn_{\rm DM}}{dt}+3Hn_{\rm DM} &\approx &=-\frac{1}{2}\langle \sigma v \rangle_{\chi \chi^* \rightarrow Z' Z'}n_{\rm DM}^2 + 2\langle \sigma v \rangle_{Z'Z' \rightarrow \chi \chi^*}(n_{\rm Z'}^{\rm eq})^2 \nonumber \\
&&-\frac{1}{2}\langle \sigma v \rangle_{\chi \chi \rightarrow Z' \chi^*}n_{\rm DM}^2 + \langle \sigma v \rangle_{Z'\chi^* \rightarrow \chi \chi}n_{\rm Z'}^{\rm eq} n_{\rm DM}
\eea

It can be rewritten with (4) and (5). Then, the relic density is determined to be
\bea
\Omega_{\rm DM}h^2=5.20\times10^{-10}{\rm GeV}^{-2}\bigg( \frac{g_*}{10.75} \bigg)^{-1/2} \bigg( \frac{x_f}{20} \bigg) e^{\Delta_{Z'}x_f}g(\Delta_{Z'},x_f)
\eea and the $g(\Delta_{Z'},x_f)$ is written in the original paper \cite{orig}.

We put $\langle \sigma v \rangle_{Z'Z' \rightarrow \chi \chi^*}=a$ and $\langle \sigma v \rangle_{Z'\chi^* \rightarrow \chi \chi}=bv^2=6b/x$. Due to $e^{\Delta_{Z'}x_f}$, the $2\rightarrow2$ annihilation cross section can be large, and the self-scattering cross section of dark matter also can be large.

\subsection{SIMP processes}
\label{simp}

$Z'$ and dark Higgs are heavier than dark matter and they have small couplings to dark matter and SM particles, the $3\rightarrow 2$ annihilation process for dark matter becomes dominant. 

Then, the Boltzmann equation is approximated to 
 \bea
 \frac{dn_{\rm DM}}{dt}+3Hn_{\rm DM} \approx - \langle \sigma v^2 \rangle_{3\rightarrow 2}(n_{\rm DM}^3-n_{\rm DM}^2n_{\rm DM}^{\rm eq}).
 \eea Here, the effective $3\rightarrow2$ annihilation cross section is obtained as
\bea
\langle \sigma v^2 \rangle_{3\rightarrow 2}=\frac{1}{4}\big( \langle \sigma v^2 \rangle_{\chi \chi \chi^* \rightarrow \chi^* \chi^*}+ \langle \sigma v^2 \rangle_{\chi \chi \chi \rightarrow \chi \chi^*} \big) \equiv \frac{\alpha_{\rm eff}^3}{m_{\chi}^5}.
\eea As a result, the dark matter relic density is given by
\bea
\Omega_{\rm DM}h^2 = 1.41\times 10^{-8}{\rm GeV}^{-2}\bigg( \frac{g_*}{10.75} \bigg)^{-3/4} \bigg( \frac{x_f}{20} \bigg)^2\Bigg( \frac{\alpha_{\rm eff}}{M_{P}^{1/3}m_\chi}  \Bigg)^{-3/2}.
\eea
From the result, the relic density depends on the ratio, $m_\chi/\alpha_{\rm eff}$. The $3\rightarrow 2$ self-annihilation cross sections are directly related with the self-scattering cross section, $\sigma_{\rm self}\sim \alpha_{\rm eff}^2/m_\chi^2$. Therefore, the self-scattering cross section can be large enough.

\begin{figure}
  \begin{center}
   \includegraphics[height=5cm]{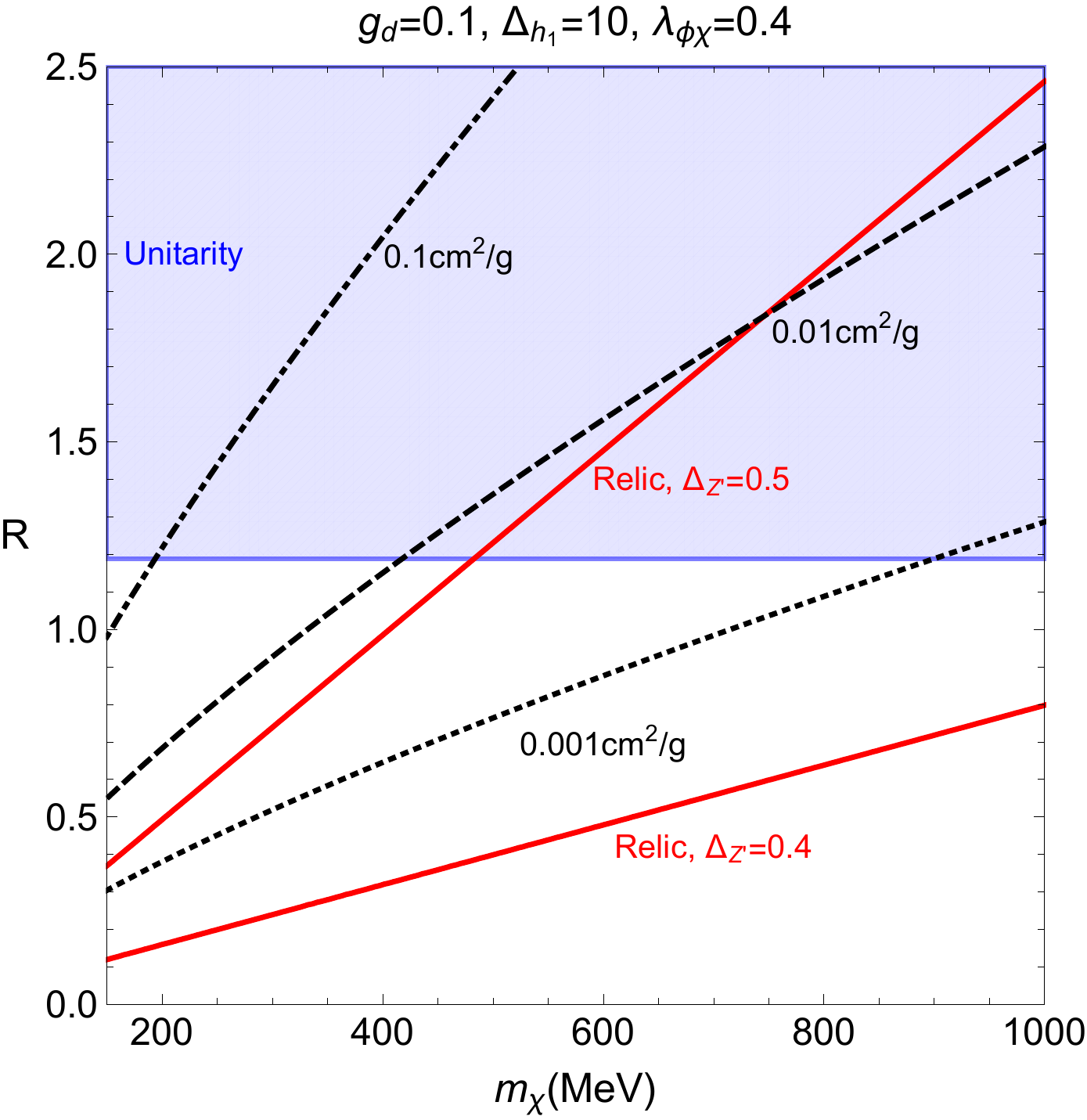}
    \includegraphics[height=5cm]{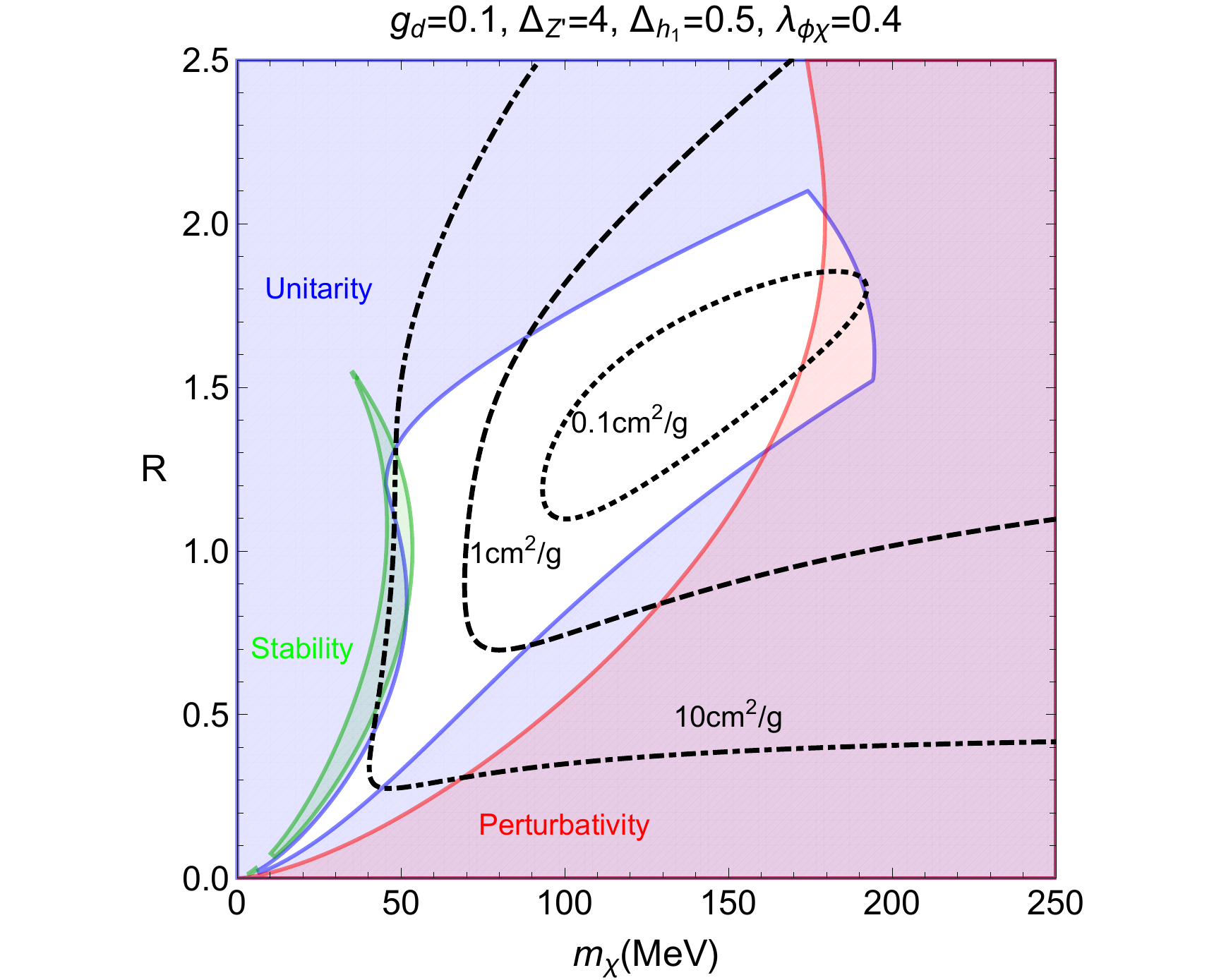} 
   \end{center}
\caption{Parameter space of $R$ vs $m_\chi$, satisfying the relic density for forbidden channel (on left) and for SIMP process (on right) with $R=\sqrt{2}\kappa v_d/6m_\chi$. The black lines correspond to self-scattering cross sections, $\sigma_{\rm self}/m_\chi=0.001, 0.01$ and $0.1\ {\rm cm^2/g}$ for left, $0.1,1$ and $10\ {\rm cm^2/g}$ for right.}
\label{fig-1}       
\end{figure}
Especially, in SIMP process, we define the effective $2\rightarrow2$ annihilation cross section by $\langle \sigma v \rangle_{{\rm eff},2\rightarrow 2}\equiv n_{\rm DM} \langle \sigma v^2 \rangle_{3\rightarrow 2}$. Then, we can rewrite as below
\bea
\langle \sigma v \rangle_{{\rm eff},2\rightarrow 2} \approx g(x/2\pi)^3 T^3 e^{-x}\langle \sigma v^2 \rangle_{3\rightarrow 2}.
\eea
In Fig.1, we note that SIMP process (on right) requires larger dark matter self-scattering cross section than the case with forbidden channels (on left). Besides, the Boltzmann supression factor in (10) is comparable with those in (4) and (5) for the forbidden channels. From this, we can infer that the forbidden channels can be dominant until $\Delta_{Z'} \sim 0.5$ and we check it exactly in the right side of Fig. 2 which is in the case with $m_\chi < m_{Z'} \ll m_{h_1}$.

\begin{figure}
  \begin{center}
   \includegraphics[height=5cm]{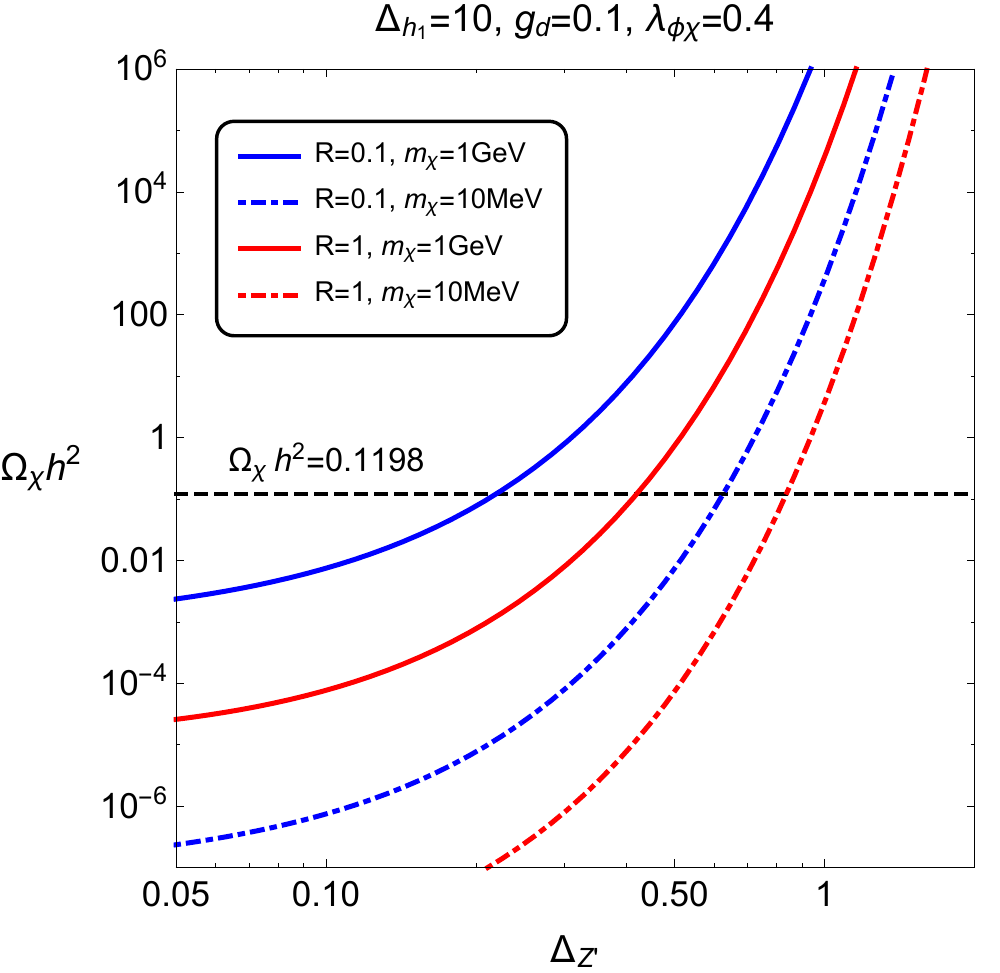}
    \includegraphics[height=5cm]{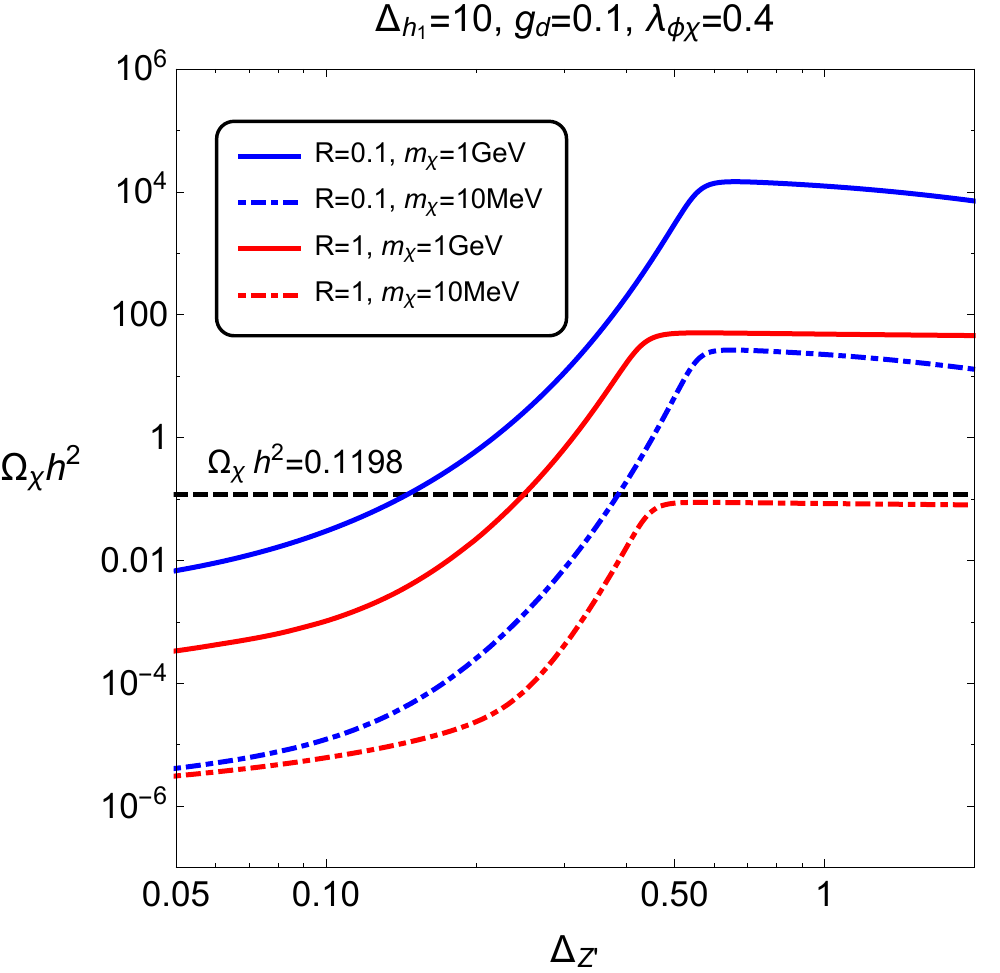} 
   \end{center}
\caption{Dark matter relic density as a function of $\Delta_{Z'}$ for only forbidden channels (on left) for both SIMP and forbidden channels (on right). The central value of relic density, $\Omega_{\chi}h^2=0.1198$, is given by black dashed lines. We have taken $R=0.1$ and $R=1.0$ with $R=\sqrt{2}\kappa v_d/6m_\chi$.}
\label{fig-1}       
\end{figure}

\section{Conclusion}
\label{con}
We have considered the thermal production from freeze-out for self-interacting dark matter in the gauged $Z_3$ model with dark photon and dark Higgs. We discussed that the forbidden channels ($\chi\chi^*\rightarrow Z' Z'$ and/or $h_1 h_1$, $\chi\chi \rightarrow Z'\chi^*$ and/or $h_1 \chi^*$), $3\rightarrow 2$ self-annihilation and the Standard Model $2\rightarrow 2$ annihilation can contribute to determining the relic density. We found that forbidden channels can produce light dark matter with sizable self-interation of dark matter to explain the small-scale problem. But the SIMP process favors lighter dark matter and larger self-interaction than forbidden channels. We identified the relevant thermal production mechanisms for self-interacting dark matter.

\section*{Acknowledgments}
The work is supported in part by Basic Science Research Program through the National Research Foundation of Korea (NRF) funded by the Ministry of Education, Science and Technology (NRF-2016R1A2B4008759). 
The work of SMC is supported in part by TJ Park Science Fellowship of POSCO TJ Park Foundation.
The work of YJK was supported by IBS under the project code, IBS-R018-D1.

%
%
%

\end{document}